\documentclass[11pt,twoside]{book}
\usepackage{vshalo}
\usepackage{longtable}
\usepackage{amsmath,amssymb}
\usepackage{graphicx}
\usepackage{lscape}
\usepackage{natbib}

\begin{document}

\pagestyle{myheadings}
\setcounter{equation}{0}\setcounter{figure}{0}\setcounter{footnote}{0}\setcounter{section}{0}\setcounter{table}{0}\setcounter{page}{1}
\markboth{Sterken, Samus \& Szabados}{VS-Halo Papers}
\title{The Luminosities of Type II Cepheids and RR Lyrae Variables}
\author{Michael Feast}
\affil{South African Astronomical Observatory and University of Cape Town} 

\begin{abstract} 
Recent work on the luminosities of type II Cepheids (CephIIs) and RR Lyrae variables is reviewed.
In the near infrared ($JHK_{s}$)
the CephIIs in globular clusters show a narrow, linear, period-luminosity relation
over their whole period range ($\sim$ 1 to 100 days). The 
CephIIs in the general field of the LMC follow this relation for periods shorter than $\sim$ 20 days.
At longer period (the region of the RV Tau stars), the LMC field stars have a significant scatter
and in the mean are more luminous than the PL relation. The OGLEIII optical data for the LMC field variables
show similar trends. Infrared colours of stars in the RV Tau period range show marked mean differences
between three groupings;
the Galactic field, the LMC field, and globular clusters. In the case of the Galactic field, at least,
this may be strongly influenced by selection effects. In the period range $\sim$ 4 to 20 days (the W Vir range)
there are stars lying above the PL relation which may be recognized by their light curves and are all likely to be
binaries. The bright Galactic variable, $\kappa$ Pav probably belongs to this group. There is evidence that
CephIIs in the general field (LMC and Galaxy) have a wider range of masses than those in globular clusters.
At present the CephII PL zero-point depends on the pulsation parallaxes of two stars. 

Zero-points of RR Lyrae $M_{V}$-[Fe/H] and $K_{s}-\log P$ relations can be obtained from trigonometrical,
statistical and pulsation parallaxes. These zero-points are compared with those for CephIIs and with the
classical Cepheid scale using variables of these three types in the LMC. Within the uncertainties ($\sim 0.1$mag)
the various scales are in agreement.  
\end{abstract}

\section{Introduction}
The RR Lyrae variables have long been considered important distance indicators for old populations
including globular clusters. On the other hand, type II Cepheids (CephIIs) have not generally be thought
of as useful distance indicators until relatively recently. The aim of the present paper is to
survey the current situation for these two classes of variables and to look briefly at the future prospects
for further improving their luminosity calibration. 
\medskip

\section{The Type II Cepheids}
 CephIIs have periods in the same range as classical Cepheids (i.e. 1 to $\sim$100 days)
but are old, low mass objects. They are found in globular clusters, the Galactic Halo and also
in the old (thick) disc. They can generally be distinguished from classical Cepheids by their
light curve shapes (e.g. Sterken \& Jaschek 1996). CephIIs are often divided into three groups 
according to their periods.
These are, BL Her stars (BL) with periods less than 4 days, W Vir stars (WV) with periods in the range
 4 to 20 days  and RV Tau stars (RV) with periods longer than 20 days. Many of the CephIIs in the RV period
range show alternating deep and shallow minima in their light curves and this usually taken as a defining
characteristic of RV stars though there is evidence (Zsoldos 1998) that globular cluster variables in this
period range do not show this effect. In the present paper the periods used for RV stars are the intervals
between successive maxima.

The fact that some globular clusters contain CephIIs with a range of periods shows that, unlike classical
Cepheids, they do not exhibit a period-mass relation. The generally accepted interpretation of these stars is that
of Gingold (1985). In this, the BL stars are evolving from a blue horizontal branch, through the instability
strip to the lower AGB. WV stars are brighter and on blue loops from the AGB and RV stars
are leaving the top of the AGB. There is some uncertainty as to whether the conventional period divisions
correspond exactly with these three phases. Sandage \& Tammann (2006) review and summarize a system of classification
based on light curve shapes that seem to be related to population characteristics and are partially correlated
with metallicities.

The metallicities of CephIIs cover a wide range, from $[Fe/H] = \sim -2.0$ in some globular clusters to
$\sim 0$ in some field stars. There is a range of differential abundances at least
in the field stars in this group. For instance
some of the stars are carbon rich.  Whilst some of these differential abundances can be understood,
at least qualitatively, in terms of dredge-up mechanisms and effects due to gas/dust separation  
some have not been properly explained.
An example is the overabundance of sodium in some BL stars but not WV stars (which are expected to
be the evolutionary products of BL stars) (see Maas et al 2007 and references there).
There is also the general problem of understanding both the metal-rich BL stars and metal-rich 
RR Lyrae variables in the general field, since they are both believed to belong to a population
without a blue horizontal-branch.

\section{Period-Luminosity Relations for Type II Cepheids}
  Early work (see references in e.g. Pritzl et al. 2003) suggested a period-luminosity (PL) relation
at optical wavelengths. however the nature of such relations, their dispersion and possible dependence on
metallicity remained uncertain. Recent studies have clarified the position considerably and it is now
possible, for instance, to see rather clearly the potentials and limitations of using CephIIs as distance indicators.
Matsunaga et al. (2006) obtained $JHK_{s}$ observations of 46 CephIIs in 26 globular clusters. The relative
distances of the clusters were obtained using RR Lyrae or HB stars and a relation of the form
\begin{equation}
M_{HB} = 0.22[Fe/H] + \gamma
\end{equation}
 PL relations with very small scatter were obtained
extending over the whole period range of these stars ($\sim$ 1 to 100 days); the scatter at $K_{s}$ is 0.14mag 
which includes 
errors
in the relative distances of the clusters. This together with the fact that there is a range of periods in some 
clusters suggests that any metal
dependence of the PL zero-point must be small. 

The OGLEIII survey of the LMC (Soszy\'{n}ski et al. 2008) provided valuable optical data on the CephIIs there.
In the period-luminosity plane, there is considerable scatter in $V$ and $I$. However in the quantity 
$W_{I} =I-1.55(V-I)$ which compensates for differential reddening and also probably for a real spread in
colour at a given period, the scatter is small (0.10 mag) for BL and WV stars which give a linear relation. The
scatter is wider for the, longer period, RV stars and in the mean these stars fall above a linear
extrapolation of the relation for the BL and WV stars. $JHK_{s}$ observations of the OGLEIII LMC CephIIs
(Matsunaga et al. 2009)
show a narrow PL of the same slope as that found in globular clusters  
for the BL and WV stars but with, again, the RVs falling in the mean above this relation and 
with the longer period stars showing a wider scatter. The RV stars are discussed further below. 

It is, of course, possible to obtain a zero point for CephII PL relations with an adopted distance modulus
for the LMC or a globular cluster scale. However, it is desirable to establish a scale for these stars
independent of other distance indicators. At present this is only possible using the pulsation
parallaxes of the two Galactic field CephIIs, V553 Cen ($\log P = 0.314$, [Fe/H] = +0.24) and
SW Tau ($\log P =0.200$, [Fe/H] = + 0.22) (Feast et al. 2008). In view of the fact that these two calibrators
are of short period, high metallicity and are also both carbon rich, it is of interest to note that the scale
implied by this calibration agrees well with other scales (see below).
Using these two calibrators one finds (Matsunaga et al. 2009);
\begin{equation}
M_{W_{I}} = -2.521(\log P -1.2) -4.12
\end{equation}
and
\begin{equation}
M_{K_{s}} = -2.410(\log P -1.2) -3.90
\end{equation}
where the slope at $W_{I}$ is from the LMC and that at $K_{s}$ from the globular clusters (with which
the LMC agrees). The uncertainty in these zero-points is $\sim 0.1$mag. Pending further work on
zero-point calibrators these equations are taken as applicable to globular cluster CephIIs over the
whole period range and to field stars of periods shorter than $\sim$ 20 days.

\section{The peculiar W Vir stars}
Feast et al. (2008) gave pulsation parallaxes for three CephIIs. Only two of them were used in the
calibration of the last section. The third star $\kappa$ Pav ($\log P = 0.959$, [Fe/H] = 0.0) was initially quite a puzzle.
This star has long been considered likely to be the nearest of the WV type stars and a prime
candidate for trigonometrical and pulsation parallax measurements. In fact, whilst the calibration
of the last section is in good accord with other scales (see section 8), the pulsation parallax result
for $\kappa$ Pav, $M_{K} = -3.77 \pm 0.07$ is 0.47mag brighter than eq 3 and taken alone would
give a quite discordant distance scale. In addition to this, though the revised Hipparcos data
is not sufficiently accurate to add much weight to the pulsation result, it does suggest that the
star is a close binary.

This puzzle appears to be solved by the OGLEIII LMC data which shows that there are a number of stars,
which Soszy\'{n}ski et al. (2008) call peculiar W Vir (pW) stars which lie above the PL relations just
discussed but below the classical Cepheid PL relations. Some of these stars show eclipses and
Soszy\'{n}ski et al. suggest they are all binaries. They also note that these stars have
distinctive light curves. An examination of the Hipparcos light curve of $\kappa$ Pav
(ESA 1997) shows it belong to this class. Like the LMC pW stars it is also too blue for its period.
Further work is required on this class of stars. One obvious possibility is that
they are binaries which have undergone mass exchange.

\section{The RV Tau stars}
As pointed out in section 3 the CephIIs in globular clusters show a linear PL, at least in the near infrared,
from $\sim$ 1 to 100 days. In the LMC however the variables of period greater than $\sim$ 20 days (the RV stars)
lie, in the mean, above this relation. There is considerable scatter amongst these stars with some of them lying
near the globular cluster line. Further work is require to see how stars differ as a function of position
with respect to the extrapolated PL relation. Period - frequency histograms (Matsunaga et al. 2009 fig 5) suggest
that the RV stars in the LMC field form a group distinct from the WV stars whereas this is not so in the
case of the Galactic globular cluster variables where the stars in the RV Tau period range seem to form
the tail of a distribution from shorter periods (The WV range). It should be noted
that for both the globular cluster sample and the LMC field, the stars with periods less than $\sim$ 4 days
(the BL stars) seem to form a separate group in a period-frequency histogram.

$(J-H)$ versus $(H-K_{s})$ plots (fig 3d of Matsunaga et al. 2009)
show marked difference between the positions of the globular cluster variables and the LMC field variables. 
Galactic field RV stars are quite distinct in such a plot with many showing near infrared excesses.  These 
stars appear to be a rather heterogeneous group and it is known that some of these stars are in quite complex
binary systems (e.g. Gielen et al. 2007 and references there). 
It should be borne in mind that the distances of such systems are not directly
known. Thus their relation to the LMC field variables in the OGLEIII sample and more particularly to the
globular cluster variables is uncertain.

\section{The Scatter in PL relations and the mass range of type II Cepheids}
It was mentioned in section 3 that the $V$ and $I$ PL relations in the LMC field 
for CephIIs with periods less than $\sim$ 20 days were wider
than in $W_{I}$ probably due to the existence of a real PLC relation. The globular clusters
NGC6388 and NGC6441 are the only ones for which satisfactory PL relations of CephIIs have been
established at optical wavelengths (Pritzl et al. 2003). The scatter in these relations
is remarkably small in $BVI$ as can be seen from table 1. In this table the results for
$BVI$ under "globulars" are from these two clusters (Feast et al. 2008) whereas the result at $K_{s}$ is
for all clusters (Matsunaga et al. 2006) and is an upper limit since it contains
the uncertainties in the relative distances of the clusters.
The other columns contain the results for the two pulsation parallax stars used in section 3.
The estimated uncertainty of each of the pulsation parallax results is 0.08 mag, but in any event
what is clear is that the deviation of these two Galactic field stars from PL relations
decreases drastically with increasing wavelength. 
A similar result applies for LMC field stars.
As in the case of classical Cepheids, this is
naturally explained by the existence of an instability strip of finite width so that stars of different mass
will have the same period at different luminosities. The effect decrease with increasing wavelength of observation.
In the case of a globular cluster the evolved stars will have only a small range of masses.
Thus these stars all lie close to a single evolutionary track passing through the instability strip
and this 
explains the very narrow optical relations for the clusters. Evidently the spread in masses amongst the
field CephIIs in the Galaxy and the LMC is larger than in globular clusters.
The alternative would be be that the PL spread in the field is due to metallicity effects.
However, V553 Cen and SW Tau have closely the same metallicities and this seems to
rule out a metallicity range as a major reason for the PL scatter.






\begin{table}[!ht] 
\caption{Scatter about Type II Cepheid PL Relations}
\smallskip
\begin{center}
\begin{tabular}{cccc}
\tableline
\noalign {\smallskip} 
& Globulars & V553 Cen & SW Tau\\ 
\noalign{\smallskip}
\tableline
\noalign{\smallskip}
B & 0.01 & +0.56 & --0.21\\ 
V & 0.07 & +0.26 & -0.25\\ 
I & 0.06 & +0.09 & --0.11\\
$K_{s}$ &  $<$0.14 & --0.04 & +0.03\\ 
\noalign{\smallskip}
\tableline
\noalign{\smallskip}
\end{tabular}  \end{center}

\end{table}

\section{The RR Lyrae stars}
It is usual to express the visual absolute magnitudes of RR Lyrae stars as;
\begin{equation}
M_{V} = \alpha ([Fe/H]+1.5) + \beta
\end{equation}
There has been much debate on the best value of $\alpha$ or, indeed, if the relation
is linear. Possibly the best empirical determination comes from RR Lyraes in the LMC field,
for which Gratton et al. (2004) found $\alpha = 0.21 \pm 0.05$. This value will be adopted
but it should be borne in mind that $\alpha$ may differ in different locations. For instance Clementini et al. (2005)
found $\alpha = 0.09 \pm 0.03$ in the Sculptor dwarf spheroidal galaxy. 
Some uncertainty does of course arise from any uncertainty in measured [Fe/H].

The value of $\beta$ can be found from trigonometrical, statistical  and pulsation parallaxes. RR Lyrae itself
is the only member of the class which has a trigonometrical parallax which is useful on its own. Results from
the HST and (revised) Hipparcos yield $M_{V}=+0.54$ for this star
which has [Fe/H] = --1.39 (Feast et al. 2008). 
The formal error of this is 0.11 mag. However it should be noted that in globular
clusters the horizontal branch  
is $\sim 0.3$mag wide at this metallicity (Sandage 1990) and also Catelan \& Cort\'{e}s (2008) find,
on the basis of Stromgren photometry,
that the stars is 0.06 mag brighter than the average RR Lyrae star of this metallicity. The real uncertainty of this
absolute magnitude is thus likely to be at least 0.15mag. 

The most elaborate work on statistical parallaxes of RR Lyraes
is that of Popowski \& Gould ( 1998 and references there).  They find $M_{V} = +0.77 \pm 0.13$ at [Fe/H] = --1.6.
The results from pulsation parallaxes depend on the models adopted (see e.g. Cacciari \& Clementini 2003). Fernley
et al. (1998)  quote a value which is equivalent to $\beta = +0.73 \pm 0.14$.
Giving equal weight to the above three values leads to $\beta = +0.68 \pm \sim 0.10$.

An infrared RR Lyrae $K -\log P$ relation was found in globular clusters by Longmore et al. (1986). A very clear
example of this is shown by the LMC cluster Reticulum (Dall'Ora et al. 2004).
This relation may be written;
\begin{equation}
M_{K_{s}} = \gamma \log P + \delta [Fe/H] + \phi
\end{equation}
Sollima et al. (2006) found $\gamma = -2.38 \pm 0.04$ from globular clusters. The metallicity term is still
very uncertain. Theoretical work summarized by Sollima et al suggests $\delta \sim 0.2$. 
They also report that observational data,
which depend on adopted relative distances of globular clusters give $\delta = 0.08 \pm 0.11$. This can 
be interpreted as either agreement with theory or a metallicity independent relation. For RR Lyrae
stars in the general field, mean period decreases with increasing metallicity (e.g. Smith 1995 fig 1.5). 
Thus to a first approximation
any metallicity dependence may be incorporated in the $\log P$ term.  The trigonometrical parallax of
RR Lyrae leads to $M_{K_{s}} = -0.64$. Jones et al (1992) found $\gamma = -2.33 \pm 0.20$ and 
$\phi = -0.88 \pm 0.06$ from pulsation parallaxes and Dambis (2009) derived $\phi = -0.82 \pm 0.08$  from
statistical parallaxes adopting the same value of $\gamma$. Both these authors assume $\delta$ to be
zero. These three results lead, with $\gamma = -2.33$, to a mean value of $\phi  = -0.97$ the uncertainty is
$\sim 0.1$mag. The trigonometrical value deviated from this mean by 0.25mag which is a little disturbing
though perhaps not statistically significant since the quoted uncertainties of results from the various
methods are generally internal values. 

An important caveat, at least when using RR Lyraes to determine the distance of globular clusters, is the fact that
some relatively metal rich clusters (e.g. NGC6441 [Fe/H] = --0.5) contain variables of this type which are
overluminous in $M_{V}$ for their metallicities (e.g Pritzl et al. 2003, Matsunaga et al. 2009). This
anomaly is connected to the ``second parameter effect"  i.e. that, unusually for relatively metal rich clusters,
these have extended horizontal branches.

\section{A Comparison of RR Lyrae, Type II Cepheid and Classical Cepheid Scales}
  A test of the consistency of the RR Lyrae and CephII scales with each other and with a classical Cepheid
scale can be made using the distances they each imply for the LMC. This is show in table 2.
\begin{table}[!ht]
\caption{Estimates of the LMC Distance Modulus}
\smallskip
\begin{center}
\begin{tabular}{ll}
\tableline
\noalign{\smallskip}
Method & Modulus\\
\noalign{\smallskip}
\tableline
\noalign{\smallskip}
Cepheids ($VI$) uncorr. & $18.52 \pm 0.03$\\
Cepheids ($K$) uncorr. & $18.47 \pm 0.03$\\
Cepheids ($VI$) corr. & $18.39 \pm 0.05$\\
CephII ($VI$) & $18.46 \pm \sim 0.1$\\
CephII ($K$) & $18.50 \pm \sim 0.1$\\
RR Lyrae ($V$) & $18.38 \pm \sim 0.1$\\
RR Lyrae ($K$) & $18.37 \pm \sim 0.1$\\
\noalign{\smallskip}
\tableline
\noalign{\smallskip}
\end{tabular}
\end{center}
\end{table}

The classical Cepheid results are based on
HST and revised Hipparcos trigonometrical parallaxes together with a period-luminosity-colour (reddening-free)
relation in $V,I$ and a period-luminosity relation in $K$ (see van Leeuwen et al. 2007 and references there).
The first two entries have not been corrected for metallicity differences between the LMC Cepheids and
the Galactic Cepheid calibrators. The third entry shows the $VI$ result after applying a metallicity correction
based on the work of Macri et al. (2006) and others (see van Leeuwen et al. 2007). 
Observational estimates of metallicity effects at $K$ are not available.

The CephII results are based
on the calibrations discussed in section 3 together with OGLE ($VI$) and IRSF ($JHK_{s}$) LMC data discussed
by Matsunaga et al. (2009). No metallicity correction has been applied (see section 3). The RR Lyrae results
depend on the zero-point calibrations of section 7. The LMC $V$ photometry and [Fe/H] values are from Gratton (2004)  
and the LMC $K$ data from Szewczyk et al. (2008).
The various estimates evidently agree well.

\section{Future work}
A programme, led by Fritz Benedict is currently in progress using the HST and ground based telescopes 
to measure the trigonometrical parallaxes of
four Galactic RR Lyrae stars. It is hoped that in the mean the distance scale error due to parallax errors
will be reduced to $\sim 0.05$mag in the modulus.  The programme includes the CephII stars VY Pyx and $\kappa$ Pav.
It is hope that the former will provide a valuable zero-point for CephII relations and that the parallax of the
latter will test the hypothesis (see section 4) that this star is brighter than normal CephIIs of its period.


\begin{thebibliography}{}      
\bibitem[]{} Cacciari, C. \& Clementini G. 2003, in, Alloin, D. \& Gieren, W. (eds)
Stellar candles for the extragalactic distance scale, Springer, Berlin, 105
\bibitem[]{} Catelan, M \& Cort\'{e}s, C, 2008, \apj, 676, L135
\bibitem[]{} Clementini, G. et al. 2005, MNRAS, 363, 734
\bibitem[]{} Dall'Ora, M. et al. 2004, \apj, 610, 269
\bibitem[]{} Dambis, A. K. 2009, MNRAS, 396, 553
\bibitem[]{} ESA, 1997, The Hipparcos Catalogue, ESA SP-1200
\bibitem[]{} Feast, M. W. et al. 2008, MNRAS, 386, 2115
\bibitem[]{} Fernley, J. et al. 1998, A\&A, 330, 515
\bibitem[]{} Gielen, C. et al. 2007, A\&A, 475, 629
\bibitem[]{} Gingold, R. 1985, Mem Soc Ast It. 56, 169
\bibitem[]{} Gratton, R. G. et al. 2004, A\&A, 421, 937
\bibitem[]{} Jones, R. V. et al. 1992, \apj, 386, 646
\bibitem[]{} Longmore, A. J., Fernley, J. A. \& Jameson, R. F. 1986, MNRAS, 220, 279
\bibitem[]{} Maas, T, Giridhar, S \& Lambert, D. L. 2007, \apj, 666, 378
\bibitem[]{} Macri, L. M. et al. 2006, \apj, 652, 1133
\bibitem[]{} Matsunaga, N. et al, 2006, MNRAS, 370, 1979
\bibitem[]{} Matsunaga, N, Feast, M. W. \& Menzies, J. W. 2009, MNRAS, 397, 933
\bibitem[]{} Pritzl, B. J. et al. 2003, \aj, 126, 1381
\bibitem[]{} Sandage, A. 1990, \apj, 350, 603
\bibitem[]{} Sandage, A. \& Tammann, G. A. 2006, ARAA, 44, 93
\bibitem[]{} Smith, H. A. 1995, RR Lyrae stars, Cambridge University Press
\bibitem[]{} Sollima, A., Cacciari, C. \& Valenti, E. 2006, MNRAS, 372, 1675
\bibitem[]{} Soszy\'{n}ski, I. et al. 2008, Acta Ast. 58, 293
\bibitem[]{} Sterken, C. \& Jaschek, C, 1996, Light curves of variable stars, Cambridge University Press.
\bibitem[]{} Szewczyk, O. et al. 2008, \aj, 136, 272
\bibitem[]{} van Leeuwen, F. et al. 2007, MNRAS, 379, 723
\bibitem[]{} Zsoldos, E. 1998, Acta Astron. 48, 775




\end{thebibliography}
\end{document}